\documentclass[preprintnumbers,amsmath,amssymb]{revtex4}
\usepackage{graphics}
\usepackage{dcolumn}

\begin{document}

\title{Einstein gravity with torsion induced by the scalar field }
\author{H. T. \"{O}z\c{c}elik*, R. Kaya*, M. Horta\c{c}su**}
\address{*Y$\i$ld$\i$z Technical University, Physics Department, 34220 Davutpa\c sa, Istanbul, TURKEY.
\\** Mimar Sinan Fine Arts University, Physics Department, 34380 Bomonti, Istanbul, TURKEY}

\begin{abstract}
We couple a conformal scalar field in (2+1) dimensions
to Einstein gravity with torsion. The field equations are obtained by a
variational principle.  We could not solve the Einstein and Cartan
equations analytically. These equations are solved numerically
with 4th order Runge-Kutta method. From the numerical solution, we
make an ansatz for the rotation parameter in the proposed metric,
which gives an analytical solution for the scalar field for
asymptotic regions.
\end{abstract}

\maketitle

\section{\label{sec:level1}Introduction}

In a series of three recent papers \cite{Sur1,Sur2,Sur3} dark
energy is explained in terms of metric-scalar couplings with
torsion. In these papers, Sur and Bhatia discuss ''the replacement
of the cosmological constant $\Lambda$ with a scalar field
non-minimally coupled to curvature and torsion to overcome the
problems of the cosmological constant due to the ''fine tuning''
problem \cite{Carroll,Pad,Cline,Krauss,Witten,Bousso,Shiu} to
explain the driving of the late time cosmic acceleration
\cite{Sur3,Riess,Peri}''. Sur and Bhatia state that \cite{Sur3}
''although observations greatly favor the $\Lambda CDM$ model
(cold dark matter with $\Lambda$) \cite{Hinshaw,Bennett,Ade1,Ade2}
there is still some room for models which replace the cosmological
constant with scalar field coupled to torsion''. This fact
motivates researchers to investigate models which couple fields
with torsion as well as curvature.

The Einstein gravity with torsion is the simplest generalization of
Einstein's general relativity theory, allowing the possibility of
relating space-time with torsion. It reduces to Einstein's
original theory when torsion vanishes. The Einstein's general
relativity is in agreement with all experimental facts in the
domain of macrophysics. It has been argued, however, in the
microscopic level space-time must have a non-vanishing torsion,
and so, microscopic gravitational interactions should be described
by the Einstein gravity with torsion \cite{Heh2}. It has been also shown
that torsion is required for a complete theory of gravitation
\cite{Humm}. The spin of matter, as well as its mass plays a
dynamical role this theory. All the available
theoretical evidence that argues for admitting spin and torsion
into a gravitational theory is summarized in Ref.s
\cite{Von,Ham,Shap}.

The spin-gravity coupling has been paid much attention and
appeared in the work of several authors, who have been mainly
interested in the study of the matter fields, namely, scalar,
gauge, and spinor fields \cite{Saa,Dzh}.

Among the other recent papers on the arXiv for Einstein
 gravity with torsion, one can cite the paper by Ivanov and Wellenzohn
\cite{Ivan} where the torsion field acts as the origin of the
cosmological constant or dark energy density. Still another paper
treats helicity effects of solar neutrinos using a dynamic torsion
field \cite{Castillo}. Torsion is also necessary for the stability
of self-accelerating universe \cite{Daemi,Nikio}. Minkevich solves
acceleration with torsion instead of dark matter \cite{Mino}.
Torsion can also be a source for inflation \cite{Akho}. Alencar
finds that torsion is necessary to localize the fermion field in
the Randall-Sundrum 2 model \cite{Alencar}.

The non-minimally coupled scalar field is of interest for general
relativistic gravitational theories, and plays an important role
in inflationary cosmology \cite{Gal}. Exact general solutions of
the Einstein and Cartan equations for open Friedmann models containing
a non-minimally coupled scalar field with an arbitrary coupling
constant have been obtained \cite{Galik}. Galiakhmetov continued
working on this field and wrote several papers where the scalar
field coupled to torsion and curvature gave rise to interesting
results \cite{Gal1,Gal2,Gal3,Gal4,Gal5}.

Studying models in lesser dimensions to disclose some properties
of similar models in (3+1) dimensions has been a common method in
quantum field theory. General relativity in (2+1) dimensions has
become an increasingly popular endeavor to understand the basic
features of the gravitational dynamics \cite{Car}.

The study of (2+1) dimensional gravity led to a number of
outstanding results, among which the discovery of the Ba\~{n}ados,
Teitelboim and Zanelli (BTZ) black hole is of particular
importance \cite{Btz,Btzm}. Einstein gravity in (2+1) dimensions
coupled to a scalar field is studied in the literature
\cite{Mart,Henn,Has1,Hasp,Schmidt}. The interest in
Einstein gravity with torsion in (2+1) dimensions has also
grown in recent years \cite{Gar,Miel,Bla,Bla1,Bla2,Bla3}.

In this work we study the Einstein gravity with torsion in (2+1)
dimensions conformally coupled with a scalar field.  Conformally
coupling refers to the fact the matter term in the action is
invariant under conformal transformations. We should note that in the framework of Einstein gravity with torsion, a scalar field
non-minimally coupled to gravity gives rise to torsion, even
though the scalar field has zero spin. In this paper, we present
the geometrical apparatus necessary for the formulation of the
Einstein gravity with torsion. By variation of the action function with
respect to vielbein and Lorentz connection we obtain the field
equations in a general form. We attempt to solve these equations
numerically. Studying the plots of these equations, we make an
ansatz for the rotation parameter in the proposed metric ending up
in an analytical solution for the scalar field.

The work organized as follows: In Section 2 we introduce notation
and definitions used throughout this work. We set up the total
action function of the scalar field with gravitation. We give
explicit form of the Klein-Gordon equation,  Einstein and Cartan
field equations. In Section 3 we work out the Einstein and Cartan
field equations which, after simplification, reduce to the system
of first and second order differential equations. These equations
can not be solved analytically. By using the methods of 4th order
Runge-Kutta we give the numerical solutions. From our graphs, we
can conjecture the form of the angular momentum parameter $J$. For
this case we can find the asymptotic form of the scalar field
analytically. In Section 4 we conclude with some final remarks and
perspectives. In an Appendix we give some technical details.

\section{\label{sec:level1}Einstein gravity with torsion and conformal coupled scalar fields in (2+1) dimensions }

The Einstein gravity with torsion is the closest theory with
torsion to general relativity. We used the massless scalar field
as the source of torsion.

We take a  homogenous and isotropic universe, i.e. we assume that our
solutions will be functions of only the radial coordinate $r$, a circularly symmetric solution.
The line element (2+1) dimensional space-time is given by
\begin{eqnarray}
ds^2 = -(v(r)+\frac{J(r)^2}{r^2})dt^2+w(r)^2dr^2+(r
d\phi+\frac{J(r)}{r}dt)^2
\end{eqnarray}
in plane polar coordinates $(t,\,r,\,\phi)$. This metric describes an AdS black hole with $g_{tt}=-v(r)$ and  $g_{rr}=w^2(r)$.  Through solving $ w^{-2}(r_h)=0$, we can obtain the radius of the black hole event horizon. Here $J(r)$ is the angular momentum parameter.

We consider a massless scalar field non-minimally coupled to
Einstein gravity with torsion in (2+1) space-time dimensions in the
presence of a cosmological constant $\Lambda$.

The action for the gravitational field with torsion is given by
\begin{eqnarray}
S=\int\sqrt{-g}L d^3x.
\end{eqnarray}
The Lagrangian for Einstein gravity  with torsion and with a massless scalar
field can be written as:
\begin{eqnarray}
L=L_\mathcal{G}+ L_\mathcal{M}+L_\mathcal{I}.
\end{eqnarray}
Here $L_\mathcal{G}$ represents the Lagrangian of the
gravitational field
\begin{eqnarray}
L_\mathcal{G} = \frac{1}{2\kappa}(R-2\Lambda),
\end{eqnarray}
$L_\mathcal{M}$ represents the Lagrangian of the matter field
\begin{eqnarray}
L_  \mathcal {M} =-\frac{1}{2}
\nabla^{\mu}\varphi\nabla_{\mu}\varphi,
\end{eqnarray}
and $L_\mathcal{I}$ represents the interaction between the
gravitational field and the matter field
\begin{eqnarray}
L_  \mathcal{I} =-\frac{1}{2} \xi R\varphi^2
\end{eqnarray}
with (2+1) space-time metric tensor $g_{\mu\nu}$, the determinant
of the metric tensor $g$, the scalar field $\varphi$, the Einstein
gravitational constant $\kappa$, the non-minimally coupling
constant $\xi$ and the Ricci scalar of the Riemann-Cartan
space-time $R$.

The Greek indices $(\mu,\,\nu, \,\sigma)$ refer to the space-time,
and they run over $1,\,2,\,3$. From physical arguments
Riemann-Cartan space-time is assumed to possess a connection
\begin{eqnarray}
\Gamma^\rho_{\mu\nu}=\{ ^{\rho}_{
\mu\nu}\}-{K_{\mu\nu}}^\rho.
\end{eqnarray}
$\{ ^{\rho}_{ \mu\nu}\}$ are Christoffel symbols of the second
kind which are symmetric in their covariant indices built up from
the metric tensor $g_{\mu\nu}$
\begin{eqnarray}
 \{ ^{\rho}_{
 \mu\nu}\}=\frac{1}{2}g^{\rho\sigma}(\partial _ \mu g_{\nu\sigma}+
 \partial _ \nu g_{\mu\sigma}- \partial _ \sigma g_{\mu\nu}),
\end{eqnarray}
and already appearing in Einstein's relativity.

As in Einstein gravity with torsion, the anti-symmetric part of the
connection $\Gamma^\rho_{\mu\nu}$ defines Cartan's torsion tensor,
\begin{eqnarray}
{T_{\mu\nu}}^\rho=(\Gamma^\rho_{\mu\nu}-\Gamma^\rho_{\nu\mu}).
\end{eqnarray}
${K_{\mu\nu}}^\rho$ is the contortion tensor, which is given in
terms of the torsion tensor by
\begin{eqnarray}
{K_{\mu\nu}}^{\rho}=\frac{1}{2}(-{T_{\mu\nu}}^{\rho}+T_{\mu
 \,\,\,\nu}^{ \,\, \, \rho}-{T^{\rho}}_{\mu\nu}).
\end{eqnarray}
Cartan's torsion tensor ${T_{\mu\nu}}^\rho$ and the contortion
tensor ${K_{\mu\nu}}^{\rho}$, in contrast to the Einstein gravity with torsion, both vanish identically in conventional
general relativity \cite{Hehl}.

The contracted torsion tensor
\begin{eqnarray}
T_{\mu}={T_{\sigma \mu}}^\sigma
\end{eqnarray}
is the torsion trace vector \cite{Niko}. The torsion can interact
with a scalar field only through its trace. We derive this result in an Appendix, along the lines our reference  \cite{Krec}.  We also agree with the results obtained in  \cite{Hojman}.

The connection (7) is used to define the covariant derivative of a
contravariant vector,
\begin{eqnarray}
\nabla _ \nu A^{\mu}=\partial_\nu A^{\mu}+ \Gamma^\mu_{\nu\rho}
A^{\rho}.
\end{eqnarray}

The Riemann-Cartan curvature tensor is defined by using the
connection (7), and is given by
\begin{eqnarray}
 R^\rho_{\sigma\mu\nu}=\partial_\mu\Gamma^\rho_{\nu\sigma}-\partial_\nu\Gamma^\rho_{\mu\sigma}+\Gamma^\rho_{\mu\lambda}
\Gamma^\lambda_{\nu\sigma}
-\Gamma^\rho_{\nu\lambda}\Gamma^\lambda_{\mu\sigma}.
\end{eqnarray}

The Ricci tensor of the Riemann-Cartan connection is defined as
$R_{\mu \nu}= R^\rho_{\mu\rho\nu}$. The scalar curvature of the
Riemann-Cartan space-time is given as follows
\begin{eqnarray}
R=g^{\mu\nu}R_{\mu \nu}.
\end{eqnarray}
The curvature scalar $R$ can be presented in the form
$R=\tilde{R}+R(T)$, where $\tilde{R}$ is the Riemannian part of
the curvature built from the Christoffel symbols; $R(T)$ is the
part of which is obtained from the covariant derivative of the
torsion tensor \cite{Krec}.

The curvature tensor for any connection satisfies the following
identities \cite{Spi},
\begin{eqnarray}
 -\left(T^{\mu}{}_{\rho\tau; \nu} + T^{\mu}{}_{\nu \rho;\tau} +
          T^{\mu}{}_{\tau \nu; \rho}\right) + R^{\mu}{}_{\nu \rho\tau} +
    R^{\mu}{}_{\rho\tau \nu} +
    R^{\mu}{}_{\tau \nu \rho}   \nonumber \\ - \left(T^{\eta}{}_{\rho\tau} T^{\mu}{}_{\eta \nu} +
          T^{\eta}{}_{\tau \nu} T^{\mu}{}_{\eta \rho} +
          T^{\eta}{}_{\nu \rho} T^{\mu}{}_{\eta\tau}\right) = 0
\end{eqnarray}
(Bianchi's first identity).
\begin{eqnarray}
 R^{\delta}{}_{\mu \nu \rho;\tau} + R^{\delta}{}_{\mu \rho\tau;
\nu} +
    R^{\delta}{}_{\mu\tau \nu; \rho}+ \left(R^{\delta}{}_{\mu\eta \nu} T^{\eta}{}_{\rho\
\tau} + R^{\delta}{}_{\mu\eta \rho} T^{\eta}{}_{\tau \nu} +
          R^{\delta}{}_{\mu\eta\tau} T^{\eta}{}_{\nu \rho}\right) = 0
\end{eqnarray}
(Bianchi's second identity). We find that the Bianchi identities
are satisfied identically.

In an orthonormal frame the metric tensor of space-time can be
expressed as follows
\begin{eqnarray}
g_{\mu \nu}= {e_{\mu}}^a {e_{\nu}}^b \eta_{a b},
\end{eqnarray}
where $\eta_{a b}=(-,+,+)$ is (2+1) Minkowski metric and
${e_{\mu}}^a $ is the vielbein field.

We can also take the metric with the co-tetrad fields
\begin{eqnarray}
ds^2 = -(e^1)^2+(e^2)^2+(e^3)^2.
\end{eqnarray}
By comparing  the above metric and the metric (1), the co-tetrad fields can be obtained as
\begin{eqnarray}
&& {e}^1=\frac{\sqrt{J^2+r^2v}}{r} dt,
\nonumber \\& &{e}^2=w dr,
\nonumber \\
& & {e}^3= r d\phi+ \frac{J}{r} dt.
\end {eqnarray}
From the expression ${e}^a={e_a}^\mu d x^\mu$,  an orthonormal base for the metric (1) can be found
\begin{eqnarray}
{e_a}^\mu=\left(\begin{array}{ccc}
\frac{\sqrt{J^2+r^2v}}{r}&0&0\\0&w&0\\ \frac{J}{r}&0&r
\end{array}\right).
\end{eqnarray}

A covariant derivative of the covariant metric tensor vanishes,
\begin{eqnarray}
\nabla _{\lambda}g_{\mu \nu} = {\partial_\lambda}g_{\mu \nu} -
      g_{\eta \nu}\Gamma^{\eta}{}_{\lambda\mu} -
      g_{\mu  \eta}\Gamma^{\eta}{}_{\lambda \nu} = 0.
\end{eqnarray}

This condition is referred to as metricity or metric compatibility
of the affine connection.

The $\Gamma$-connection may be introduced by imposing the vielbein
postulate
\begin{eqnarray}
\nabla _{\mu}{e_{\nu}}^a = {\partial_{\mu}} {e_{\nu}}^a
+{e_{\nu}}^b {\omega_{\mu}{}^a{}_b} -
      {e_{\rho}}^a \Gamma^{\rho}{}_{\mu \nu}=0.
\end{eqnarray}
From the vielbein postulate we solve the $\Gamma$-connection as
follows
\begin{eqnarray}
{\Gamma^{\rho}}_{\mu \nu} = {e_{a }}^{\rho}({\partial_\mu
{e_{\nu}}^{ a }} +{e_{\nu}}^{b }{\omega_{\mu}{}^a{}_b} ).
\end{eqnarray}

\subsection{ Explicit form of the Klein-Gordon equation and Cartan equations}

The Klein-Gordon equation in an external gravitational field with
torsion is considered. By Hamilton's principle, the variation of
the total action $S$ for the gravitational field, matter field and
interaction of torsion with matter field vanishes $\delta S=0$.

By varying the total action with respect to the scalar field
$\varphi$
\begin{eqnarray}
\frac{\partial(\sqrt{-g} L)}{\partial
\varphi}-\partial_{\rho}\frac{\partial(\sqrt{-g}
L)}{\partial(\partial_{\rho}\varphi)}=0,
\end{eqnarray}
the Klein-Gordon equation obtained as follows:
\begin{eqnarray}
&& -16 \kappa \xi^2 w \varphi \varphi'^2(J^2 +
          r^2 v)^2 + (J^2 +
          r^2 v)(\kappa\xi (8\xi + 1)\varphi^2 -
          1){}
              \nonumber \\&& (\kappa\xi\varphi^2 -
          1)(2 w\varphi''(J^2 +
              r^2 v)
              + \varphi'(-2 w'(J^2 + r^2 v)  +
              r w(r v' +
                    2 v)))
                     {}
              \nonumber \\&& +\xi\varphi(\kappa\xi\varphi^2 \ - 1)^2(2(J^2 + r^2 v)(2 J w
J'' + r^2 w v'
 - 2 r v w'){}
              \nonumber \\&& - 2 r v'(r w'(J^2 + r^2 v)
            -  w(3 J^2 + r^2 v)) + 2 J w J'+ 4 J^2 v w -
        4 J J' {}
              \nonumber \\&&(w'(J^2 + r^2 v)
       + r w(r v' + 2 v)) -
        w J'^2(J^2 - 3 r^2 v) -
        r^4 w v'^2)=0.
\end{eqnarray}

Here $'$ denotes the derivative with respect to $r$.

Varying the total action with respect to the Lorentz connection
field ${\omega _{\mu}}^{a b}$
\begin{eqnarray}
\frac{\partial(\sqrt{-g} L)}{\partial {\omega _{\mu}}^{a b}} -
\partial_{\rho}\frac{\partial(\sqrt{-g}
L)}{\partial(\partial_\rho {\omega _{\mu}}^{a  b})} = 0,
\end{eqnarray}
gives Cartan field equations. From the equation (23) obtained from the vielbein postulate and equation (26), we can obtain the following Cartan field equations;
\begin{eqnarray}
&&
 r\left({\omega _2}^{31}\sqrt{J^2 + r^2 v} - J' + r w{\omega
_1}^{32}\right) +
    J = 0,\; \; \; {}
    \nonumber \\
    &&
 J' + J{\omega _2}^{33} + r w\left({\omega _1}^{23} - {\omega
        _1}^{32}\right) =
    0,\; \; \; {}
    \nonumber \\
    &&
 r\left({\omega _2}^{31}\left(-\sqrt{J^2 + r^2 v}\right) + J' +
          r w{\omega _1}^{23}\right) - J = 0, \; \; \; {}
    \nonumber \\
    &&
 J' + J U + r w\left({\omega _1}^{23} - {\omega _1}^{32}\right)
          = 0,\; \; \;{\omega _2}^{11} +U=0,\; \; \; {\omega _2}^{33} - U=0, {}
\nonumber \\
&&
J^2\left(r\left(U + {\omega _2}^{11}\right) - 1\right) + J
r\left(-{\omega _2}^{31}\sqrt{J^2 + r^2 v}+ J' -
r w{\omega _1}^{32}\right) {}
              \nonumber \\&&+ r^3 v\left(U + {\omega _2}^{11}\right) =
0,\; \; \;{}
\nonumber \\
&&
 r\left({\omega _2}^{31}\left(-\sqrt{J^2 + r^2 v}\right) + J' +
          r w{\omega _1}^{23}\right) + J\left(r{\omega _2}^{33} - 1\right) =
          0,\; \; \;{}
\nonumber \\
&&
J\left(1 - r{\omega _2}^{33}\right) -
    r\left({\omega _2}^{31}\left(-\sqrt{J^2 + r^2 v}\right) + J' +
          r w{\omega _1}^{23}\right) = 0, \; \; \;{}
\nonumber \\
&&{\omega _2}^{33} = U, \; \; \;
{\omega _1}^{33} = 0, \; \; \;
{\omega _2}^{32} = 0, \; \; \;
{\omega_1}^{31} = 0, \; \; \; {\omega _1}^{22} = 0,
\end{eqnarray}
where
\begin{eqnarray}
U=\frac{2 \kappa \xi \varphi\varphi'}{\kappa \xi \varphi^2-1}.
\end{eqnarray}

Cartan field equations (27) can be solved and remaining Lorentz connection coefficients can be obtained as follows
\begin{eqnarray}
&&{\omega _1}^{12} = \frac{J J' +
        r^2 v'}{2 r w\sqrt{J^2 +
            r^2 v}}, \; \; \; {\omega _1}^{21} = -{ \omega _1}^{12}-\frac{U\sqrt{J^2 +
              r^2 v}}{r w} ,  \; \;\; {\omega _1}^{32} = \frac{J'}{2 r w},  {}
\nonumber \\
&& {\omega _1}^{23} = -{\omega _1}^{32}
              -\frac{J U}{r w}, \; \;\; {\omega _2}^{13} =-{\omega} _2^{31}= \frac{2 J - r J'}{2
r\sqrt{J^2 + r^2 v}},  \; \;\;{\omega _2}^{11} =- U,{}
\nonumber \\
&&  {\omega _3}^{12} =-{\omega _3}^{21}=
\frac{2 J - r J'}{2 w\sqrt{J^2 + r^2 v}} \; \;\; {\omega _3}^{23} = - {\omega
_3}^{32}-\frac{r U}{w}, \; \;\;  {\omega _3}^{32}
= \frac{1}{w}, {}
\nonumber \\
&& {\omega_1}^{13} = 0, \; \;  \; {\omega _1}^{11} = 0,  \; \; \;  {\omega_2}^{12} = 0, \; \; \;
{\omega_2}^{21} = 0, \; \; \;  {\omega_2}^{22} = 0, \; \;  \; {\omega
_2}^{23} = 0,  \nonumber \\
&&  {\omega _3}^{11} = 0, \; \; \;  {\omega _3}^{13}
= 0, \; \; \;   {\omega_3}^{22} = 0, \; \; \; {\omega _3}^{31} = 0, \;
\;  \;  {\omega _3}^{33} = 0.
\end{eqnarray}

From the equations (7), (23) and (29), we can arrive non-zero
components of the contortion tensor
\begin{eqnarray}
 {K}_{1 2}^1={K}_{3 2}^3 = -U,\; \;\;{K}_{1 1}^2 =
-\frac{v}{w^2}U,\; \;\; {K}_{1 3}^2 = {K}_{3 1}^2 =
\frac{J}{w^2}U,\; \;\;{K}_{3 3}^2 = \frac{r^2}{w^2}U.
\end{eqnarray}
Substituting the Lorentz connection  ${\omega _{\mu}}^{a b}$ (29) into equation (23) we can obtain the components of the $\Gamma$-connection as follows
\begin{eqnarray}
&&
\Gamma _{12}^1= \Gamma _{21}^1+U,  \; \;
  \Gamma _{21}^1=\frac{J J'+r^2 v'}{2 J^2+2 r^2 v}, \; \;   \Gamma _{11}^2=\frac{2U v+v'}{2 w^2},  \;  \;   \Gamma _{22}^2=\frac{w'}{w},
\nonumber \\
&&\Gamma _{13}^2=\Gamma _{31}^2=-\frac{J'+2J U}{2 w^2}, \; \;
 \Gamma _{12}^3= \Gamma _{21}^3=\frac{v J'-J v'}{2 J^2+2 r^2 v}, \;\; \Gamma _{33}^2=-\frac{r^2 U+r}{w^2},  \nonumber \\
&&\Gamma _{32}^3= \Gamma _{23}^3+U, \; \;    \Gamma _{23}^3=\frac{J J'+2 r v}{2 J^2+2 r^2 v},\;\;\Gamma _{23}^1= \Gamma _{32}^1=-\frac{r \left(r J'-2 J\right)}{2 \left(J^2+r^2 v\right)},
 \nonumber \\
&&   \Gamma _{11}^1=0,\; \;    \Gamma _{11}^3=0,\; \;     \Gamma _{12}^2=0, \; \;  \Gamma _{13}^1=0, \; \;   \Gamma _{13}^3=0, \; \;  \Gamma _{21}^2=0,  \; \;   \Gamma _{22}^1=0, \nonumber \\
&&   \Gamma _{22}^3=0,
 \; \;  \Gamma _{23}^2=0,   \; \;  \Gamma _{31}^1=0, \; \; \Gamma _{31}^3=0,  \; \;  \Gamma _{32}^2=0, \; \;   \Gamma _{33}^1=0, \; \;  \Gamma _{33}^3=0.
\end{eqnarray}

Using the above $\Gamma$-connection, the non-zero components of the torsion tensor (9) are defined as
\begin{eqnarray}
{T}_{1 2}^1={T}_{3 2}^3=U.
\end{eqnarray}
By means of the relation (11) the trace of the torsion can be
obtained as $T_2=2 U$.

From the line element (1), the affine connection (7), the torsion
tensor (9) and the contortion tensor (10), the Cartan field
equations are obtained in the PhD thesis prepared by Hasan Tuncay
\"{O}z\c{c}elik (YTU 2016) \cite{Tunc}.

As a check on our calculations, we vary the total action with respect to the
contortion $K^\rho_{\mu\nu}$
\begin{eqnarray}
\frac{\partial(\sqrt{-g} L)}{\partial
K^\rho_{\mu\nu}}-\partial_{\sigma}\frac{\partial(\sqrt{-g}
L)}{\partial(\partial_{\sigma}K^\rho_{\mu\nu})}=0,
\end{eqnarray}
which gives the Cartan field equations. Solving these equations, we can
arrive the same result as in equation (30) \cite{Tunc}.

\subsection{Explicit form of the Einstein field equations }
The variation of the total action with respect to the vielbein
field ${e_a}^{\mu}$
\begin{eqnarray}
&&\frac{\partial(\sqrt{-g} L)}{\partial {e_a}^{\mu}} - \partial _{
\rho}\frac{\partial(\sqrt{-g} L)}{\partial{(\partial
_\rho}{e_a}^{\mu})} +
\partial _{\sigma}
\partial_{\rho}\frac{\partial(\sqrt{-g} L)}{\partial(\partial_{\sigma}\partial_{\rho} {e_a}^{\mu})}= 0
\end{eqnarray}
or the variation of the total action with respect to the metric
tensor $g_{\mu\nu}$
\begin{eqnarray}
&&\frac{\partial (\sqrt{-g}L)}{\partial
g_{\mu\nu}}-\partial_{\rho}\frac{\partial
(\sqrt{-g}L)}{\partial(\partial_{\rho}g_{\mu\nu})}+\partial_{\sigma}\partial_{\rho}\frac{\partial(\sqrt{-g}
L)}{\partial(\partial_{\sigma}\partial_{\rho}g_{\mu\nu})}=0,
\end{eqnarray}
yields the same Einstein field equations
\begin{eqnarray}
&&4 r(J^2 + r^2 v)(\kappa \xi\varphi^2 -
          1)(\Lambda r w^3 -
        2 \kappa \xi\varphi(\varphi'(w - r w') +
              r w\varphi''))
{}
          \nonumber \\&&  + 2 \kappa r^2 w \varphi'^2(J^2 +
          r^2 v) (\kappa \xi (4\xi + 1)\varphi^2 + 4\xi -
        1)
{}
          \nonumber \\ && + (\kappa \xi\varphi^2 -
            1)^2(4 r w'(J^2 + r^2 v) -
        w r J' - 2 J)^2)=0,
\end{eqnarray}
\begin{eqnarray}
&&2 \kappa \varphi'^2(J^2 +
          r^2 v)(\kappa\xi (8\xi + 1)\varphi^2 -
        1) + (\kappa\xi\varphi^2 - 1)^2 (J'^2 +
        2 r v')+ {}
\nonumber \\ && (\kappa\xi\varphi^2 -
          1)(4 \kappa\xi\varphi\varphi'(2 J J' +
              r(r v' + 2 v)) -
        4\Lambda w^2(J^2 + r^2 v))=0,
\end{eqnarray}
\begin{eqnarray}
&&4(J^2 + r^2 v)(\kappa \xi\varphi^2 -
          1)(\kappa\xi\varphi(\varphi'(w v' - 2 v w') +
              2 v w\varphi'')- \Lambda v w^3) -{}
          \nonumber \\&&
           2 \kappa v w\varphi'^2(J^2 +
          r^2 v) (\kappa\xi (4\xi + 1)\varphi^2 + 4\xi -
        1) + (\kappa\xi\varphi^2 -
            1)^2
            (2 w v''(J^2 {}
          \nonumber \\&&+
           r^2 v) -
        2 v' w'(J^2+ r^2 v) - 2 J w J' v' + v w(J')^2 -
         r^2 wv'^2)=0,
\end{eqnarray}
\begin{eqnarray}
-2 J \kappa  w(\varphi'^2(J^2 +
          r^2 v)(\kappa\xi (4\xi + 1)\varphi^2 + 4\xi -
        1) + (\kappa\xi\varphi^2 -
            1)^2(2 w(J'' {}
          \nonumber \\
          (J^2 + r^2 v) +
              J r v')  - J w J'^2 -
        J'(2 w'(J^2 + r^2 v) +
              r w(r v' + 2 v)))+
  4 {}
          \nonumber \\
          (J^2 + r^2 v)(\kappa\xi\varphi^2 -
          1)  (\kappa\xi\varphi(\varphi'(w J' - 2 J w') +
              2 J w\varphi'') - J\Lambda w^3)=0.
\end{eqnarray}

\section{\label{sec:level1}Solutions }

In this section solutions of Einstein and Cartan equations in (2+1)
dimensional space-time is given by considering scalar field as
external source for torsion of space-time.

We conjecture that as $ r$ goes to infinity, the scalar field $\varphi(r)$ goes to zero, and when $ r$ goes to zero, the scalar field $\varphi(r)$ will be large.
We will find that by appropriate choice of our constants, we can validate this conjecture.

\subsection{Case $J(r)=0$ }

We first take $J(r)=0$, the non-rotating case. For the non-rotating black hole interacting with a scalar field we take the metric as
\begin{eqnarray}
ds^2 = -v(r)dt^2+\frac{1}{v(r)}dr^2+r^2 d\phi^2.
\end{eqnarray}

In order to express the zero-torsion and torsion parts of the
Klein-Gordon equation and the Einstein field equations, we can
write the connection (7) as follows
\begin{eqnarray}
\Gamma^\rho_{\mu\nu}=\{ ^{\rho}_{ \mu\nu}\}-\alpha
{K_{\mu\nu}}^\rho.
\end{eqnarray}
The first term in equation (7) gives zero-torsion part and the
second term gives the contribution from the torsion. When
$\alpha=0$ it means zero-torsion. When $\alpha=1$ it means that
there is a torsion. 

After a long calculation, we find an equation of motion for the scalar field given as
\begin{eqnarray}
\varphi''(r)
    - \frac{
   \alpha 4 \kappa\xi\varphi (r)\varphi'(r)^2}{\kappa\xi\varphi (r)^2 -
       1} + \frac{(2\xi - 1)\varphi'(r)^2}{2\xi\varphi (r)}=0.
\end{eqnarray}

We see that both for $\alpha =0$ and $\alpha=1$, we get  
\begin{eqnarray}
\varphi(r)
     = \frac{C_1} {\sqrt{r+C_2}}
\end{eqnarray}
if we assume $0<\kappa \xi \varphi^2<<1$ for as the dominant contribution the scalar field in the asymptotic region.
Here $C_1$ and $C_2$ are constants.

$\xi=\frac{1}{8}$ is the choice to obtain a conformal coupling of the scalar field to gravity \cite {Birrell}. For 
$\alpha=1$, there are corrections of the order of $({{r+C_2}})^{-3/2}$, which are not significant for very large values of $r$.
This solution is in agreement with those made in Horta\c{c}su, \"{O}z\c{c}elik and
Yap{\i}\c{s}kan \cite{Has1}, where the coupling to torsion was
absent. In both cases the scalar field goes to zero as $r$ goes to
infinity.

\subsubsection{Case $\alpha$=1 (with torsion) and $\kappa \xi \varphi^2>>1$ }
If we solve the equation (42), the solution can be written as
\begin{eqnarray}
\varphi(r)=\frac{C_3}{(r+C_4)^{1/6}}
\end{eqnarray}
where $C_3$ and $C_4$ are constants. This solution 
differs from the $\alpha=0$ case, and have the scalar field so large as to satisfy this inequality, $\kappa \xi \varphi^2>>1$, in the
asymptotic region, where $ r$  goes to zero.
This solution may be important when the scalar field is close to the black hole.

\subsection{Case $J(r)\neq 0$}
From the Einstein field equation (36) we find
\begin{eqnarray}
 \varphi''(r)&=&  (4 w(J^2(-\kappa \xi\varphi^2 + \Lambda r^2 w^2 +
                    1) + \Lambda r^4 v w^2) +
        r(\kappa \xi\varphi^2 - 1)  {}
\nonumber \\ &&
(4 w'(J^2 + r^2 v)
        +
              w J'(4 J -
                    r J')))/(8 \kappa \xi r^2\varphi w(J^2 +
            r^2 v))   {}
\nonumber \\ && +
(\varphi'^2(\kappa \xi (4\xi +
                    1)\varphi^2 + 4\xi -
            1))/(4\xi\varphi(\kappa \xi\varphi^2 -
            1)) {}
\nonumber \\ &&+ \varphi'(w'/w - 1/r).
\end{eqnarray}

From the Einstein field equation (37)  we find
\begin{eqnarray}
 v'(r)&=&(-2 \kappa \varphi'^2(J^2 +
              r^2 v)(\kappa\xi (8\xi + 1)\varphi^2 - 1) -
      8 \kappa\xi\varphi\varphi'(\kappa\xi\varphi^2 - 1){}
\nonumber \\ && (J J' +
            r v)
+ (\kappa\xi\varphi^2 -
              1)(4\Lambda w^2(J^2 +
                  r^2 v) + (J')^2(1 -
                  \kappa \xi\varphi^2)))/{}
\nonumber \\ && (2 r(\kappa \xi\varphi^2 -
            1) (\kappa\xi\varphi^2 + 2 \kappa \xi r\varphi\varphi' - 1)),
\end{eqnarray}
or
\begin{eqnarray}
 \varphi'(r)& =&(-2 \kappa^2\xi^2\varphi^3(2 J J' + r(r v' + 2 v))
+
      2 \kappa\xi\varphi(2 J J' +
            r(r v' +
                  2 v)) +\nonumber \\&& p \sqrt{2}
\sqrt(\kappa(\kappa\xi\varphi^2 -
                  1)(2 \kappa\xi^2\varphi^2(\kappa\xi\varphi^2 -
                        1)(2 J J' +
                        r(r v' + 2 v))^2 +\nonumber \\&&(J^2 +
                       r^2 v)
(\kappa\xi (8\xi + 1)\varphi^2 -
                        1) (4 J^2\Lambda w^2 - (\kappa\xi\varphi^2 \ - 1)(J'^2 + 2 r v')
\nonumber \\&&   + 4\Lambda r^2 v w^2))))/
(2 \kappa (J^2 +
 r^2 v)(\kappa\xi (8\xi + 1)\varphi^2 - 1)),
\end{eqnarray}
where $p$ is $\pm 1$.

Substituting $\varphi''(r)$ (45) and $v'(r)$ (46) into Einstein
field equation (39) we obtain
\begin{eqnarray}
 w'(r)&=& (w(8 J(\kappa\xi\varphi^2 -
                  1)(J^2(-\kappa\xi\varphi^2 + \Lambda r^2 w^2 +
                      1) + \Lambda r^4 v w^2) +
        \nonumber \\&&  r(4(J^2 +
                       r^2 v)
(-2 \kappa\xi\varphi\varphi'(4 J -
                              r^2 J'')(\kappa\xi\varphi^2 - 1) +
                      r J''(\kappa\xi\varphi^2 - 1)^2 \nonumber \\&& -
                      J k r \varphi'^2
(\kappa\xi (16\xi +
                                    1)\varphi^2 - 1)) +
                r^2J'^3(\kappa\xi\varphi^2 - 1)^2 -
               \nonumber \\&& 6 J r J'^2(\kappa\xi\varphi^2 - 1)^2 +
                2 J'(\kappa r\varphi'(J^2 +
                              r^2 v)(4\xi\varphi(\kappa\xi\varphi^2 - 1) +\nonumber \\&&
                            r\varphi'(\kappa \xi (16\xi + 1)\varphi^2 -
                                  1)) -2
(\kappa\xi\varphi^2 -
                              1)(J^2(-2 \kappa\xi\varphi^2 +\Lambda r^2 w^2 + 2)\nonumber \\&& +
                            r^2 v(\kappa\xi\varphi^2 + \Lambda r^2 w^2 -
                                  1))))))/
(4 \ r(J^2 + r^2 v)(r J' - 2 J)\nonumber \\&&(\kappa\xi\varphi^2 -
            1)(\kappa\xi\varphi^2 + 2 \kappa\xi r\varphi\varphi' -
            1)).
\end{eqnarray}

Substituting $\varphi''(r)$, $v'(r)$ and $w'(r)$ into the
Klein-Gordon equation (25) and the Einstein equation (38) we can
obtain
\begin{eqnarray}
 J''(r)&=& (2 J \kappa r(12\Lambda\xi^2\varphi^2 w^2(J^2 +
                    r^2 v)(\kappa\xi\varphi^2 -
                  1) + \varphi'(\varphi'(J^2 +
                          r^2 v)\nonumber \\&&((\kappa\xi\varphi^2 -
                                1)
 (\kappa\xi (8\xi + 1)\varphi^2 +
                              2\xi - 1) +
                        4 \kappa\xi^2 r\varphi\varphi') +
                  4\xi r\varphi(J^2\Lambda w^2\nonumber \\&&(\kappa\xi (6\xi +
                                      1)\varphi^2 - 1)
+
                        v((\kappa\xi\varphi^2 -
                                      1)^2 + \Lambda r^2 w^2(k\xi \
(6\xi + 1)\varphi^2\nonumber \\&& - 1)))) +
      \kappa\xi r^3\varphi J'^3\varphi'(\kappa\xi\varphi^2  -
              1)^2
-
      6 J \kappa\xi r^2\varphi J'^2\varphi'(\kappa\xi\varphi^2 -
              1)^2 \nonumber \\&&
    -  J'(\kappa r\varphi'(r\varphi'(J^2 +
                          r^2 v)((\kappa\xi\varphi^2 -
                                1)
(\kappa\xi (8\xi + 1)\varphi^2 +
                              2\xi - 1) +
                      \nonumber \\&&  4 \kappa\xi^2 r\varphi\varphi') +
                  2\xi\varphi(J^2(2\Lambda r^2 w^2(\kappa\xi (6\xi \
+ 1)\varphi^2  - 1)
- 5(\kappa\xi\varphi^2 - 1)^2)
                      \nonumber \\&&+  r^2 v((\kappa\xi\varphi^2 - 1)^2 +
                              2\Lambda r^2 w^2(\kappa\xi (6\xi +
                                        1)\varphi^2 -
                                    1)))) +
(J^2 \ + r^2 v)\nonumber \\&& (\kappa\xi\varphi^2 -
                    1)(\kappa\xi\varphi^2(-\kappa\xi\varphi^2 +
                        12\Lambda\xi r^2 w^2 + 2) -
                  1)))/\nonumber \\&&
(r(J^2 +
            r^2 v)(\kappa\xi\varphi^2 - 1)^2(\kappa\xi\varphi^2 +
          2 \kappa\xi r\varphi\varphi' - 1)).
\end{eqnarray}
From the equation (14), the curvature scalar of space-time with
torsion can be obtained as
\begin{eqnarray}
R(r)=\frac{6\Lambda w^2+\kappa \varphi'(r)^2
      }{w^2(1-\kappa\xi\varphi(r)^2 )}
\end{eqnarray}

The Ricci tensor components are given in Ref. \cite{Tunc}. In the
absence of torsion, the Ricci tensor reduces to $R(r)=6\Lambda$.

We could not solve the equations $\varphi''(r)$ (45), $v'(r)$ (46), $w'(r)$ (48) and $J''(r)$ (49) in analytical forms. We
will give only the numerical solutions using the methods of 4th
order Runge-Kutta.

We set $\kappa=1$  \cite{Bunn}, $\xi=\frac{1}{8}$, $\Lambda=10^{-8}$,
$\varphi(1)=10$, $\varphi'(1)=-2.065$, $v(1)=10^2$,
$v'(1)=\Lambda$, $w(1)=10^{-1}$, $J(1)=10^{-2}$ and
$J'(1)=10^{-4}$.

The plots of the scalar field $\varphi(r)$ and $\varphi'(r)$ with
the angular momentum $J(r)\neq0$ are given in Fig. 1. From the
Fig. 1 we can see that the scalar field goes to zero $r$ goes to
infinity.

The plots of the metric components $v(r)$ and $w(r)$ with the
angular momentum $J(r)\neq 0$ are given in Fig. 2.

The plots of the angular momentum $J(r)$ and $J'(r)$ are given in
Fig. 3.

The plot of the Ricci scalar with the angular momentum $J(r)\neq0$
is given in Fig. 4.

\begin{figure*}[!hbt]
\begin{center}$
\begin{array}{cc}
\scalebox{0.47}{\includegraphics{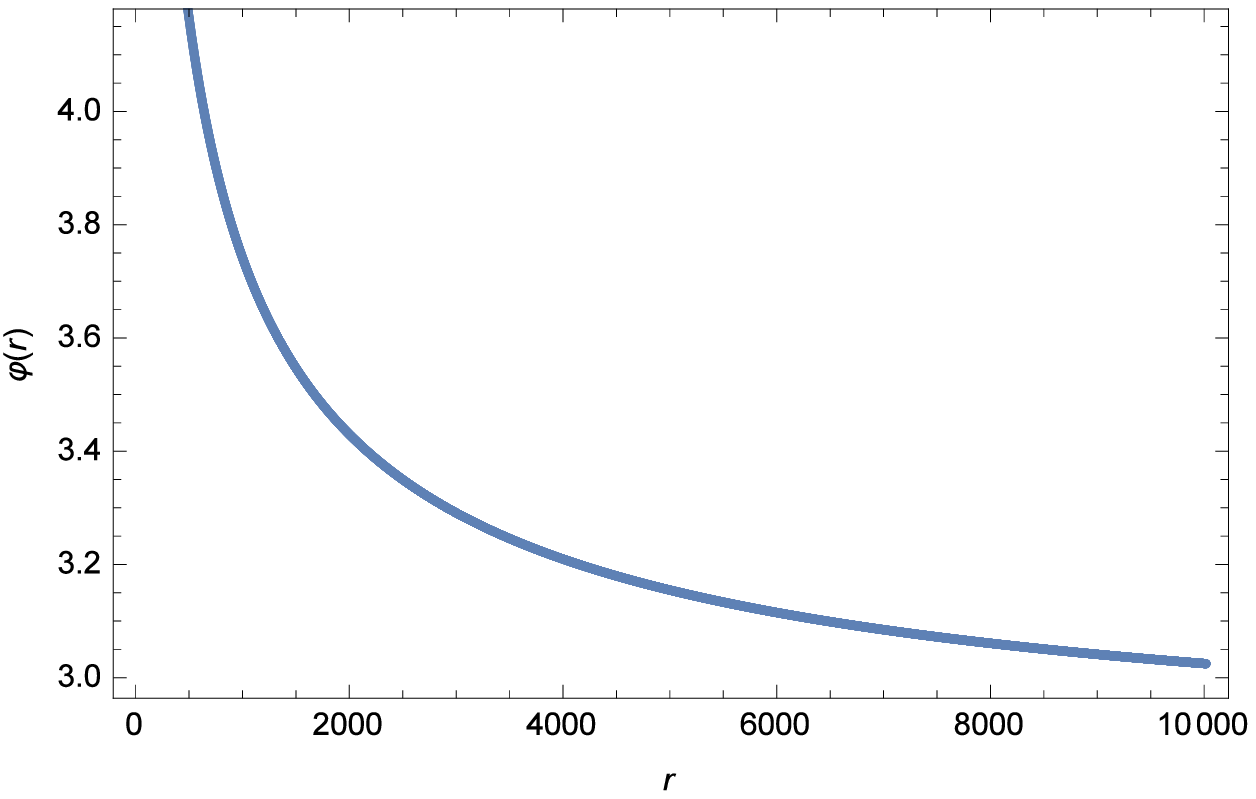}}&
\scalebox{0.47}{\includegraphics{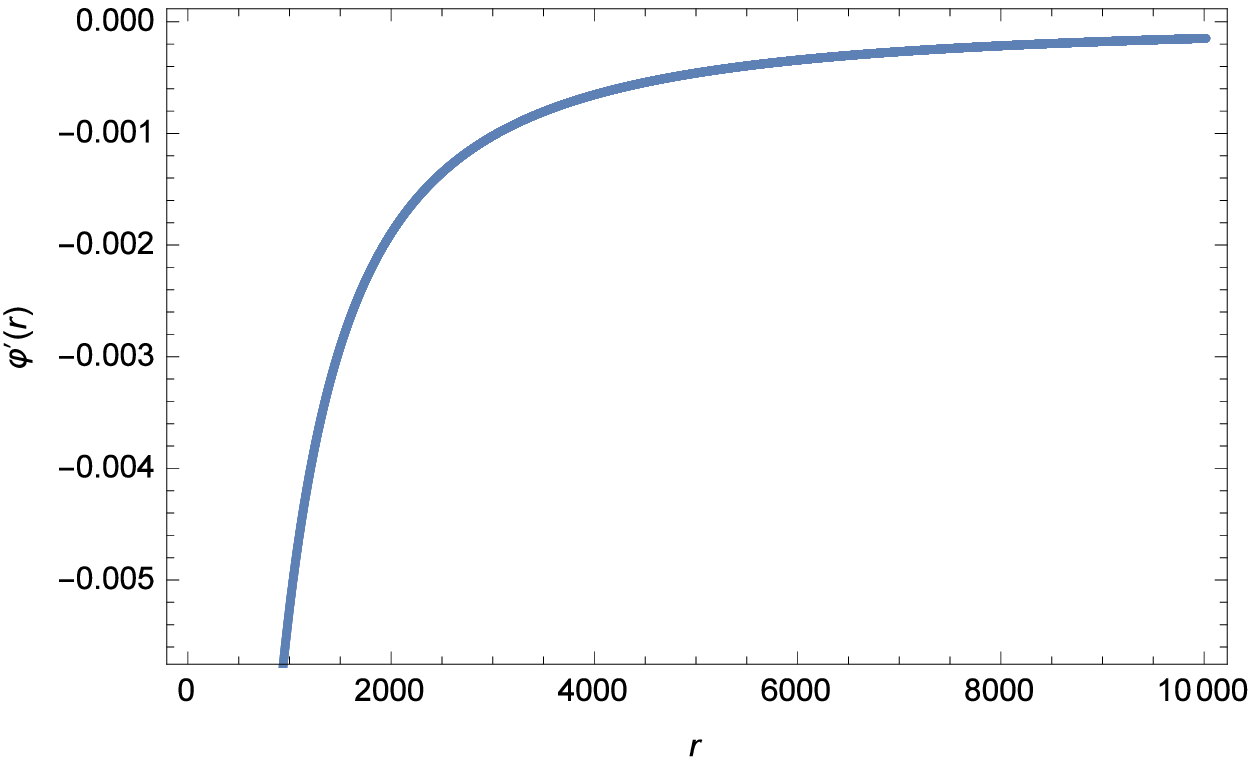}}
\end{array}$
\end{center}
\caption{The plots of the scalar field $\varphi(r)$ and
$\varphi'(r)$ with the angular momentum $J(r)\neq0$ are plotted by
Runge-Kutta method with respect to $r$. We set $\kappa=1$,
$\xi=\frac{1}{8}$, $\Lambda=10^{-8}$, $\varphi(1)=10$,
$\varphi'(1)=-2.065$, $v(1)=10^2$, $v'(1)=\Lambda$,
$w(1)=10^{-1}$, $J(1)=10^{-2}$ and $J'(1)=10^{-4}$. }
\end{figure*}
\begin{figure*}[!hbt]
\begin{center}$
\begin{array}{cc}
\scalebox{0.47}{\includegraphics{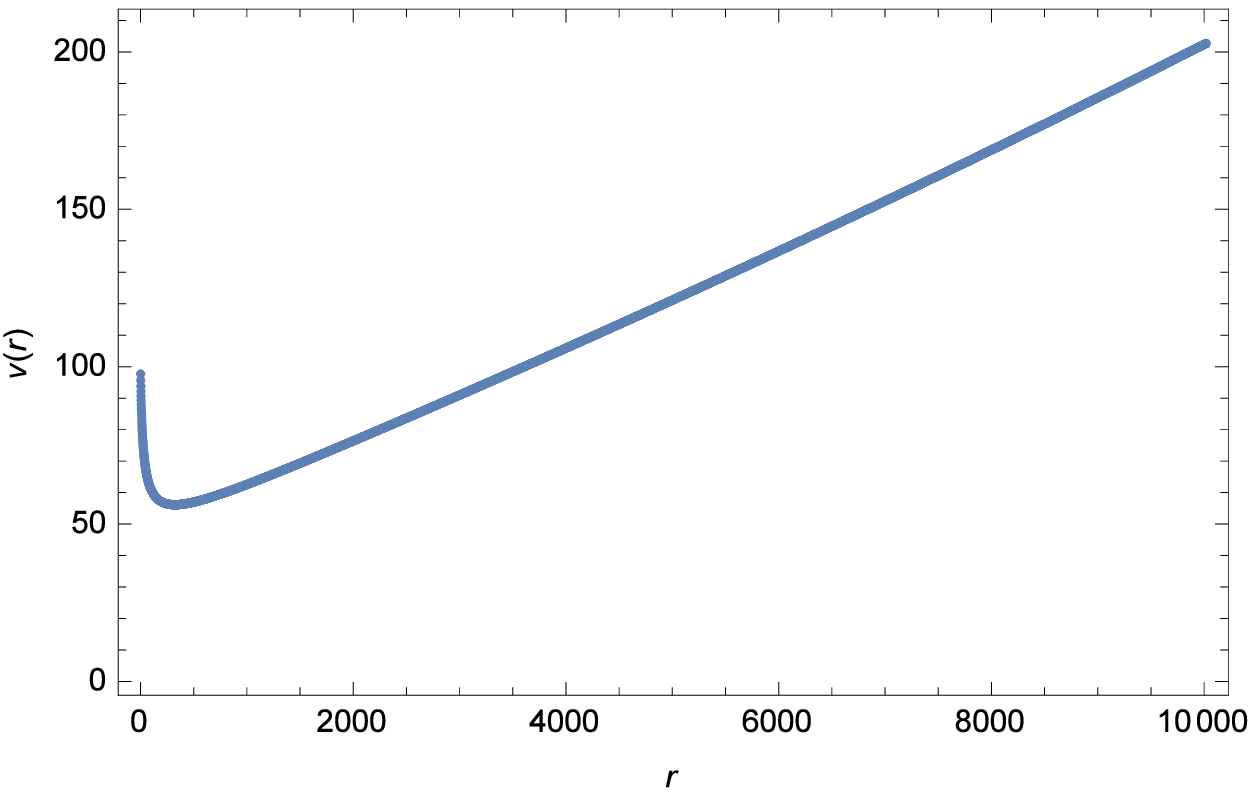}}&
\scalebox{0.47}{\includegraphics{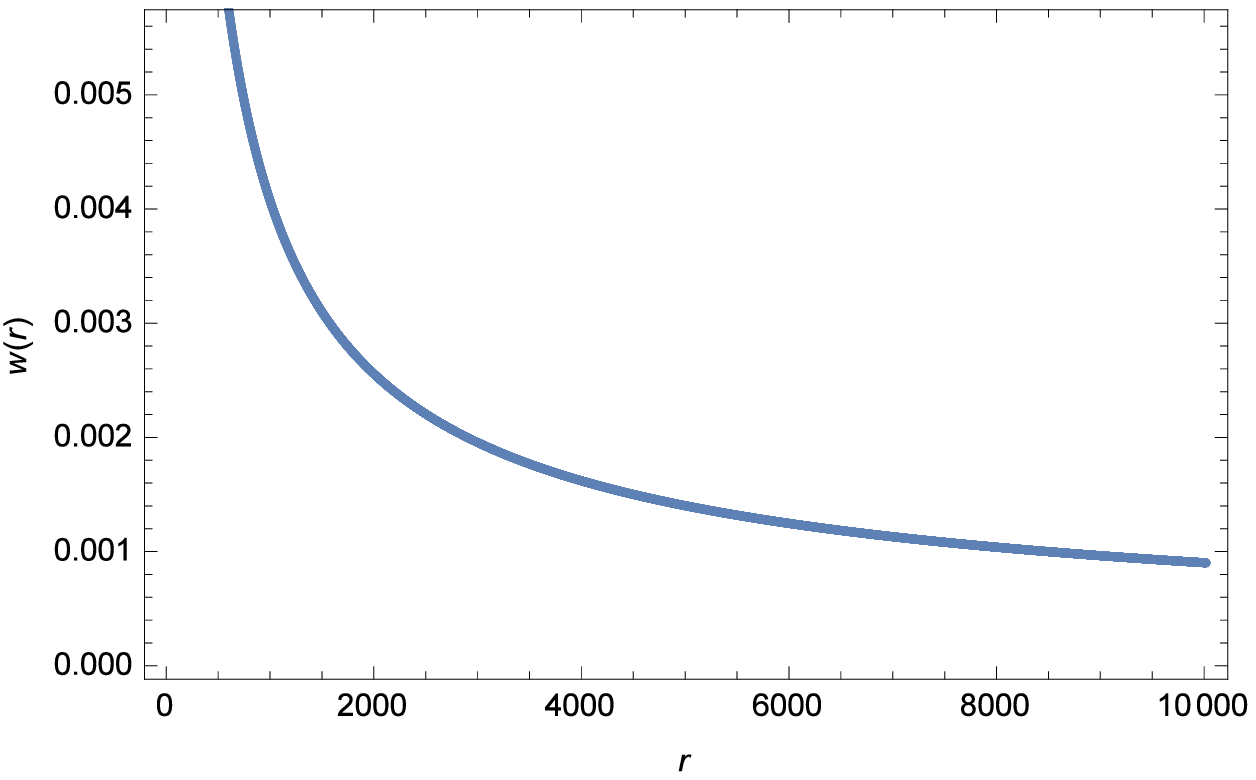}}
\end{array}$
\end{center}
\caption{The plots of the metrics components $v(r)$ and $w(r)$
with the angular momentum $J(r)\neq0$ by Runge-Kutta method with
respect to $r$. We set $\kappa=1$, $\xi=\frac{1}{8}$,
$\Lambda=10^{-8}$, $\varphi(1)=10$, $\varphi'(1)=-2.065$,
$v(1)=10^2$, $v'(1)=\Lambda$, $w(1)=10^{-1}$, $J(1)=10^{-2}$ and
$J'(1)=10^{-4}$. }
\end{figure*}
\begin{figure*}[!hbt]
\begin{center}$
\begin{array}{cc}
\scalebox{0.47}{\includegraphics{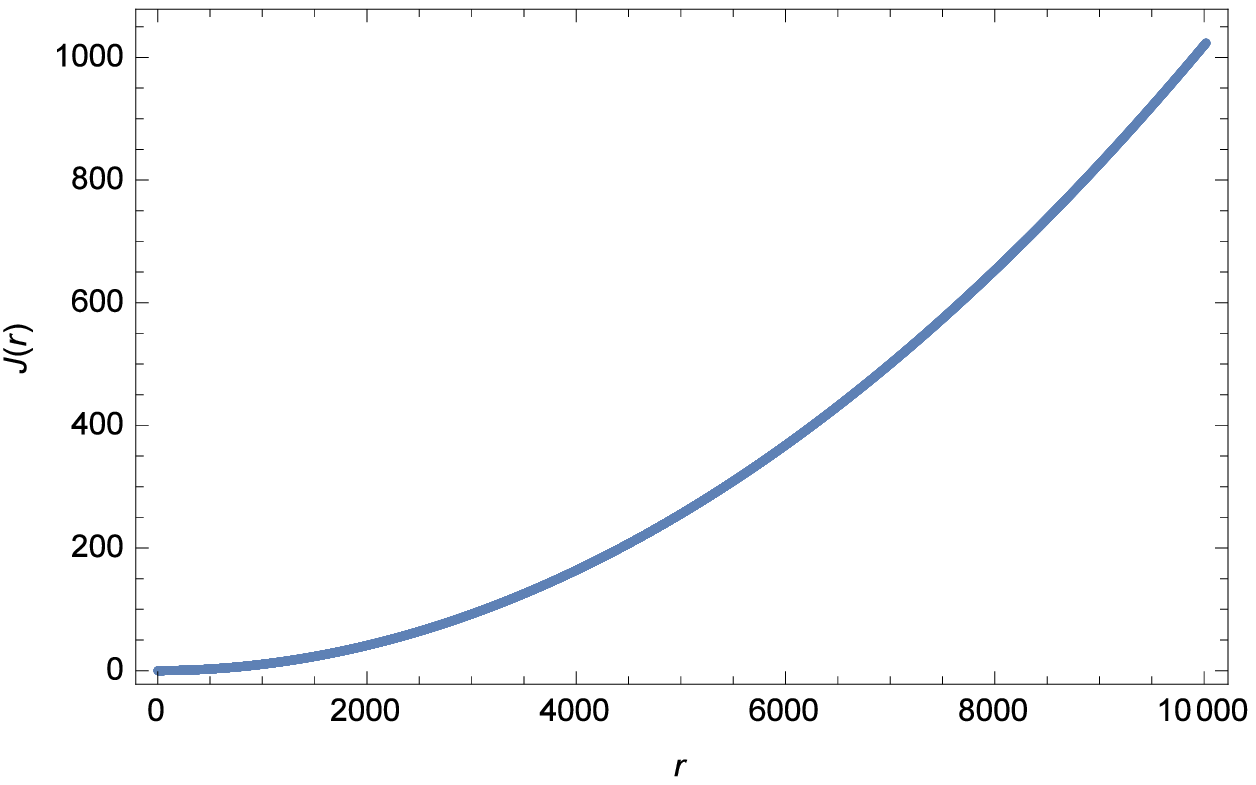}}&
\scalebox{0.47}{\includegraphics{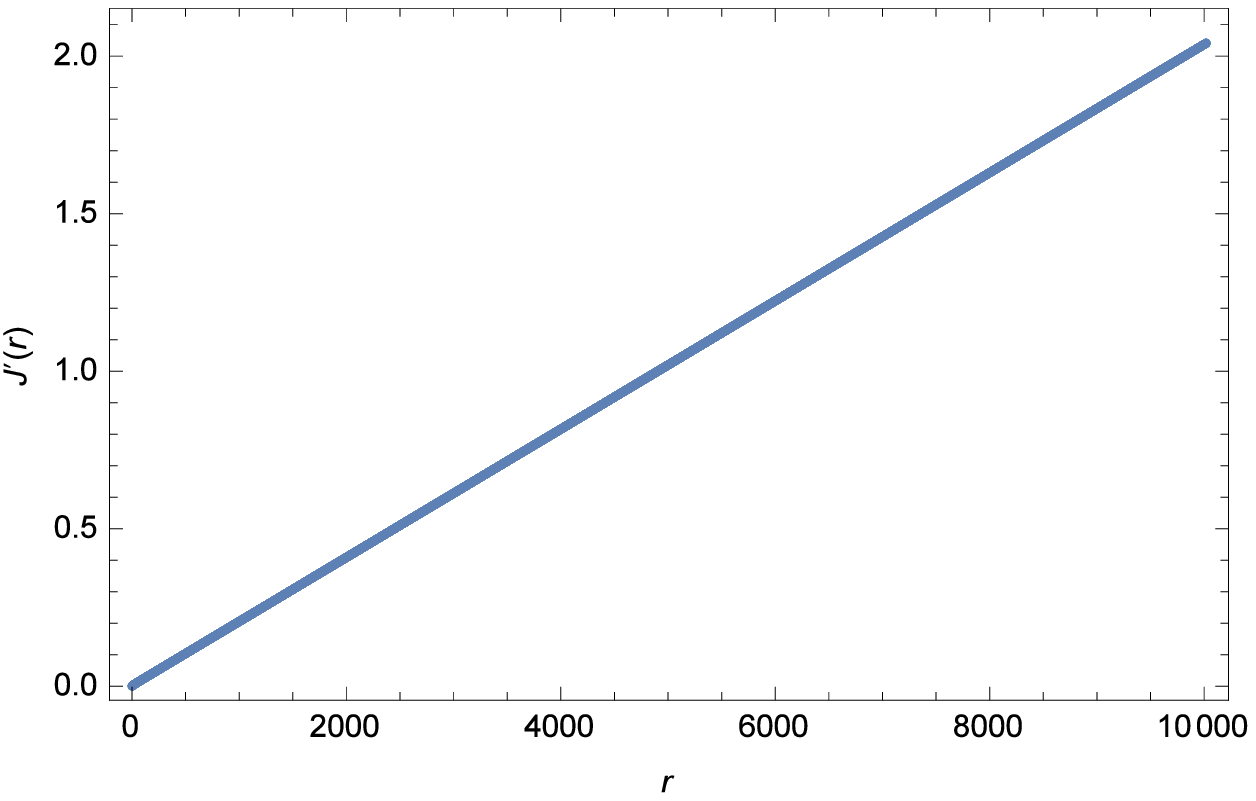}}
\end{array}$
\end{center}
\caption{The plots of the angular momentum $J(r)$ and $J'(r)$ are
plotted by Runge-Kutta method with respect to $r$. We set
$\kappa=1$, $\xi=\frac{1}{8}$, $\Lambda=10^{-8}$, $\varphi(1)=10$,
$\varphi'(1)=-2.065$, $v(1)=10^2$, $v'(1)=\Lambda$,
$w(1)=10^{-1}$, $J(1)=10^{-2}$ and $J'(1)=10^{-4}$. }
\end{figure*}
\begin{figure*}[!hbt]
\begin{center}$
\begin{array}{cc}
\scalebox{0.47}{\includegraphics{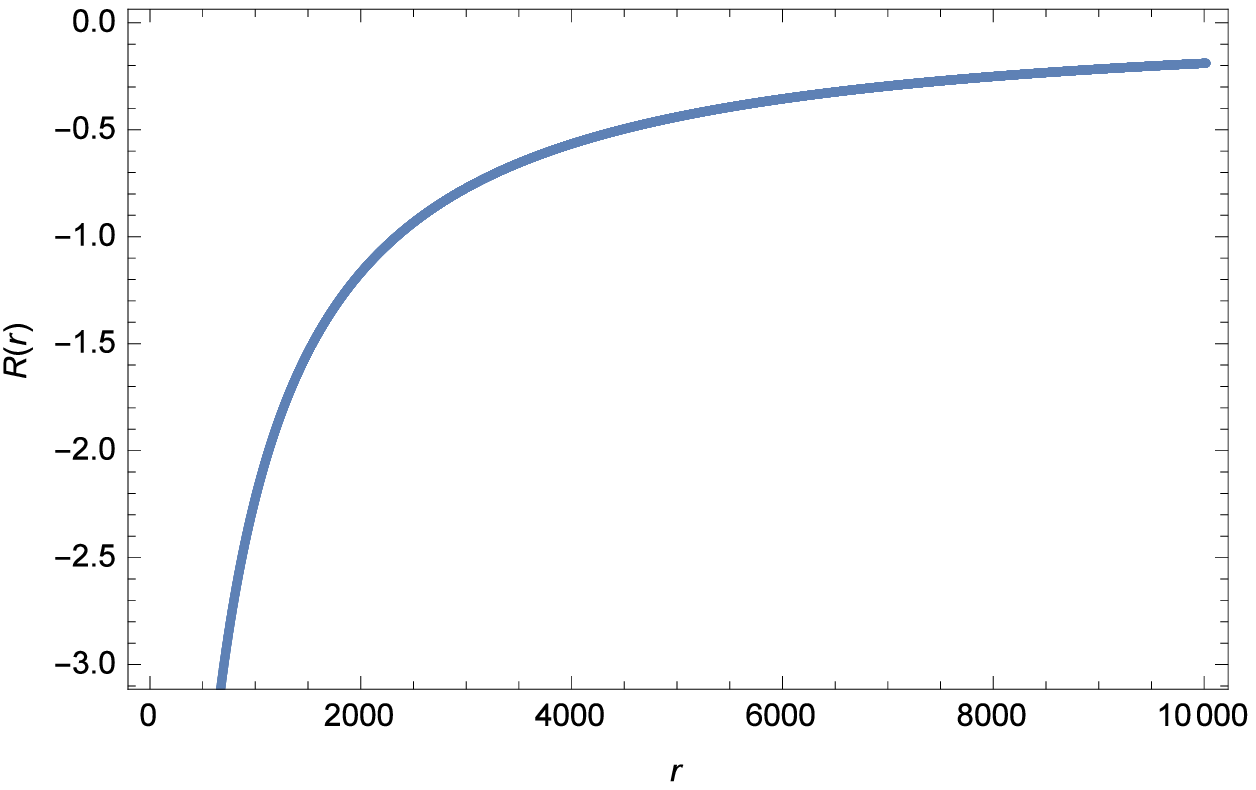}}
\end{array}$
\end{center}
\caption{The Ricci scalar $R(r)$ with the angular momentum
$J(r)\neq0$ is plotted by Runge-Kutta method with respect to $r$.
We set $\kappa=1$, $\xi=\frac{1}{8}$, $\Lambda=10^{-8}$,
$\varphi(1)=10$, $\varphi'(1)=-2.065$, $v(1)=10^2$,
$v'(1)=\Lambda$, $w(1)=10^{-1}$, $J(1)=10^{-2}$ and
$J'(1)=10^{-4}$. }
\end{figure*}

\subsection{Case $J(r)=a r^2+b$}

From the $J(r)$ and $J'(r)$ in Fig. 3, we make an ansatz for the
functional form of the angular momentum $J(r)$ as $J(r)=a r^2+b$.
If we substitute this form of $J(r)$ into $\varphi''(r)$ (45), we
can obtain
\begin{eqnarray}
\varphi''(r)-\frac{\varphi'(r)}{r}-\frac{3(3\varphi(r)^2-8)\varphi'(r)^2}{(\varphi(r)^2-8)\varphi(r)}=0.
\end{eqnarray}

\subsubsection{Case $\kappa \xi \varphi^2>>1$}

If we solve the above equation (51), the dominant part of the solution can be written
as follows
\begin{eqnarray}
\varphi(r)=\frac{C_5}{(r^2+ C_6)^{1/8}}
\end{eqnarray}
where $C_5$ and $C_6$ are constants. This case may be important when $r$ is very close to origin, 
not for  the asymptotic case $ r$ going to infinity. 
\subsubsection{Case $0<\kappa \xi \varphi^2<<1$}
From the equation (51), it is possible to express the scalar field, in the asymptotic region for very large $r$ 
as
\begin{eqnarray}
\varphi(r)=\frac{C_7}{\sqrt{r^2+C_8}}
\end{eqnarray}
where $C_7$ and $C_8$ are constants, up to terms which are of order $(r^2+C_8)^{-3/2}$. Here we see that the solution is  distinctly
different from the $J(r)=0$ case, given above.

In solving eq. (51), we made the crude approximation by taking the first relevant terms, resulting in equations (52) and (53). The essential difference between eq. (42) and eq. (51), the equation to be used when $J(r)$  is not set to be zero, is the presence of the term with first derivative of the  $\phi$ field divided by the independent variable $r$. To cancel this term, we have to use the square of $r$  as the independent variable which results in equations (52) and (53).

Our three solutions given in equations (44, 52, 53) are asymptotic solutions, in the limits $r$ going to zero or to infinity where we only take the dominant contributions. There are lower order additions to all three, which are not significant in the appropriate asymptotic values of $r$.

\section{\label{}Conclusions}

We studied in (2+1) dimensions Einstein gravity with torsion conformally
coupled to a massless scalar field $\varphi(r)$. By considering
variations with respect to the scalar field, the metric tensor and
contortion tensor, the Klein-Gordon equation, Einstein and Cartan
field equations are obtained. We plotted $J(r)$ and its derivative to have an idea of its analytical behaviour from its graph.
From the plot we made the ansatz that it is given as $J(r)=ar^2+b$. We derived solutions in
$(2+1)$ dimensions with and $J(r)=ar^2+b $ interacting with
a scalar field. We gave numerically solutions of  Einstein and Cartan
equations with $J(r)\neq 0$ in (2+1) dimensional space-time.

If the angular momentum is taken as $J(r)=a r^2+b$ and the scalar field with torsion is too small
($0<\kappa \xi \varphi^2<<1$), the scalar field $\varphi(r)$ asymptotical
behaves like $r^{-1}$. We compare it with the case without torsion, found in equation (43), which is $ r^{-1/2} $ and deduce that the
torsion has an effect on the scalar field.

With our ansatz for the angular momentum, $J(r)=a
r^2+b$, and the scalar field with torsion is large ($\kappa \xi
\varphi^2>>1$), the dependence of the scalar field $\varphi(r)$ on $(r^2+ C_6)^{-1/8}$ differs from the $J=0$, $\alpha=1$ case. When $J=0$, 
$\alpha=0$, the solution is the same as equation (43). Note that we do not expect such a behaviour for $r$ very large. It may be important if $r$ is very small. We see that both for large and small $r$, 
the angular momentum, and its source in our case, torsion, has an effect on the scalar field. We also note that to see the full effect of torsion, we have to assume $J(r)$  is not equal to zero, which necessitates a rotating metric ansatz, our equation (1).

We can conclude that the torsion has a significant effect on the scalar
field in (2+1) dimensions. The scalar field goes to zero faster in the asymptotical region, for large $r$.

 \section*{Appendix}
The torsion tensor in 2+1 dimensions has nine components. Three of
these are given by the trace, six are given by the trace-free part
of the torsion tensor, as stated in equation (3) of the paper by
V.G. Krechet and D.V. Sadovnikov \cite{Krec}. Here we use $Q$ instead of $T$.
$\widetilde{Q_{ik}^{l} }=  Q_{ik}^{l} + \frac{1}{3} ( \delta_{k}^{l} Q_{i}- \delta_{i}^{l} Q_{k} )$ .
Here $\widetilde{Q_{ik}^{l} }$ is the full torsion tensor, $ Q_{ik}^{l}$ is the trace-free part, $Q_{i}$ is the trace of the torsion tensor; $Qi = Q_{li}^{l}$. To find the degrees of freedom of torsion tensor we write the
Ricci tensor explicitly including the two different parts of
torsion, the trace-free part and the trace, equation (4) of \cite{Krec}.
Then we can argue, as Krechet et al. that the trace-free part is
not excited by our interaction, explicitly given by our lagrangian
given as in equations (3, 4, 5, 6) of our paper, making the trace-free
components null, "since the sources of it are absent in this
particular case" \cite{Krec}. The Ricci scalar is given by equation (4)
of the Krechet reference, after equating the Weil non-metrical term
$W_{i}$ to zero. $\widetilde{R} = R  +{ \beta_1} Q^{l}_{,l}   - {\beta_2 }Q^{l}Q_{l} $.
Here  $\widetilde{R}$  is the Ricci scalar with torsion, $R$  is
the part without torsion, $ { \beta_1}$ and ${ \beta_2}$ are constants. Then, by varying the action only with
respect to trace of the torsion tensor, we can find the components
of the trace of the torsion tensor in terms of the parameters in
the Lagrangian.

The trace of the torsion tensor has three components. The
resulting equations, obtained by varying the action with respect
to the trace of the torsion tensor, are given by Krechet et al.
\cite{Krec} (equation 7 d) after we equate the non-metricity term to
zero.
\begin{eqnarray}
Q_{i}={ \gamma}{ \kappa }\frac{\varphi\varphi_{,i}}{(1-\frac{\kappa}{8} \varphi^2)}
\end{eqnarray}
where ${ \gamma}$ is constant. Here ${,i} $ denotes differentiating with respect to the
coordinate $i$, since the right-hand-side of this equation is a total derivative, we may write $ Q_{i} $ as derivative of a scalar field. 
\begin{eqnarray}
Q_{i}=-\partial_{i} [\frac {\gamma}{4} \ln (1-\frac{\kappa}{8} \varphi^2)]
=\partial_{i} \Phi.
\end{eqnarray}
A similar equation is also given in the paper by J. B.
Fonseca-Neto et al, equation (6) \cite{Neto}. Both papers state that
the trace of the torsion tensor is proportional to the derivative
of the scalar field with respect to space and time coordinates. We also agree with the results obtained in \cite{Hojman}.  In
our case, the constant differs since we use a different
normalization.

In our paper, we are looking for a static solution for the scalar
field as a function of only the radial coordinate. This makes the
derivatives of the scalar field with respect to the time and
angular variable equal to zero, and leaves only the second
component of the trace of the torsion tensor non-vanishing.

We can also solve for the torsion components by varying the
action with respect to Lorentz connection field, equations (26, 27),
whose solutions are given in equation (29). We use our equations (7, 21, 29)
to construct the non-zero components of the contortion tensor, our
equation (30). We get the components of the trace of the torsion tensor
using our equation (7, 9).

\section*{Acknowledgement}  We thank the 
anonymous referee for informing us that eq. (54) 
can be written as in eq. (55) and 
several suggestions. M. H. thanks the Science Academy, Istanbul for support. 
This work is supported by T\"{U}B\.{I}TAK, the Scientific and Technological Council of Turkey.


\begin{thebibliography}{00}
\bibitem{Sur1}S. Sur  and A.S. Bhatia,  arXiv:gr-qc/1611.00654.
\bibitem{Sur2}S. Sur  and A.S.  Bhatia,  arXiv:gr-qc/1611.06902.
\bibitem{Sur3}S. Sur  and A.S.  Bhatia,  arXiv:gr-qc/1702.01267.
\bibitem{Carroll}S.M. Carroll, Living Rev. Rel. {\bf 4}, 1 (2001).
\bibitem{Pad}T.  Padmanabhan, Phys. Rept. {\bf 380}, 235 (2003).
\bibitem{Cline}J.M. Cline,  arXiv:hep-th/0612129v5.
\bibitem{Krauss} L.M. Krauss and S.J. Ray, Gen. Rel. Grav. {\bf 39}, 1454 (2007).
\bibitem{Witten} E. Witten, arXiv:hep-ph/0002297v2.
\bibitem{Bousso} R. Bousso, Pontifical Acad. Sci. Scr. Varia {\bf 119}, 129 (2014)
\bibitem{Shiu}G. Shiu and B. Greene,  { \it Perspectives on String Phenomenology} (World Scientific, Singapore, 2015).
\bibitem{Riess}A.G. Riess, et al., Astron. J. {\bf 116}, 1009 (1998).
\bibitem{Peri} S. Perlmutter, et al.,  Astrophys J. {\bf 517}, 565 (1999).
\bibitem{Hinshaw}G.F. Hinshaw, et al., Astrophys. J. Suppl. Series {\bf 208}, 19 (2013).
\bibitem{Bennett}C.L. Bennett, et al., Astrophys. J. Suppl. Series {\bf 208}, 20 (2013).
\bibitem{Ade1}P.A.R. Ade, et al., Astron. and Astrophys. {\bf 594}, A13 (2016).
\bibitem{Ade2}P.A.R. Ade, et al., Astron. and Astrophys. {\bf 594}, A14 (2016).
\bibitem{Heh2} F.W. Hehl, Found. of Phys. {\bf 15}, 451 (1985).
\bibitem{Humm}R.T. Hammond, Gen. Rel. Grav. {\bf 42}, 2345 (2010).
\bibitem{Von}F.W.  Hehl, P. von der Heyde and G.D. Kerlick, Rev. of Mod. Phys. {\bf 48}, 393 (1976).
\bibitem{Ham}R.T. Hammond, Gen. Rel. Grav. {\bf 31}, 233 (1999).
\bibitem{Shap}I.L. Shapiro, Phys. Rept. {\bf 357}, 113 (2002).
\bibitem{Saa}A. Saa, arXiv:gr-qc/9309027v1.
\bibitem{Dzh}V. Dzhunushaliev and D. Singleton, Phys. Lett. A {\bf 254}, 7 (1993).
\bibitem{Ivan}A.N. Ivanov and A. Wellenzohn, Astrophys. J. {\bf 829}, 47 (2016),
arXiv:gr-qc/1607.011128v2.
\bibitem{Castillo}D. Alvarez-Castillo, D.J. Cirilo-Lombardo and J. Zamora-Saa, arXiv:hep-ph/1611.02137.
\bibitem{Daemi}V. Nikiforova, S. Radjbar-Daemi and V. Rubakov, Phys. Rev. D {\bf 95}, 024013 (2017).
\bibitem{Nikio}V. Nikiforova, arXiv:hep-th/1705.00856.
\bibitem{Mino}A.V. Minkevich, arXiv:gr-qc/1704.06077.
\bibitem{Akho}S. Akhshabi, E. Qorani and F. Khajenabi, arXiv:gr-qc/1705.04931.
\bibitem{Alencar}G. Alencar, arXiv:hep-th/1705.09331.
\bibitem{Gal}A.M. Galiakhmetov, Russian Phys. Jour. {\bf 44}, 1316 (2001).
\bibitem{Galik}A.M. Galiakhmetov, Grav. Cosmol. {\bf 10}, 300 (2004).
\bibitem{Gal1}A.M. Galiakhmetov, Grav. Cosmol. {\bf 14}, 190 (2008).
\bibitem{Gal2}A.M. Galiakhmetov, Class. Quantum Grav. {\bf 27}, 055008 (2010).
\bibitem{Gal3}A.M. Galiakhmetov, J. Mod. Phys. D {\bf 21}, 1250001 (2012).
\bibitem{Gal4}A.M. Galiakhmetov, Gen. Rel. Grav. {\bf 44}, 1043 (2012).
\bibitem{Gal5}A.M. Galiakhmetov, Gen. Rel. Grav. {\bf 45}, 275 (2013).
\bibitem{Car} S. Carlip, Korean Phys. Soc. {\bf 28}, 447 (1995).
\bibitem{Btz}M. Ba\~{n}ados, C. Teitelboim and J. Zanelli, Phys. Rev. Lett. {\bf 69}, 1849 (1992).
\bibitem{Btzm}M. Ba\~{n}ados, M. Henneaux, C. Teitelboim and J. Zanelli, Phys. Rev. D {\bf 48}, 1506 (1993).
\bibitem{Mart}C. Mart\'{i}nez and J. Zanelli, Phys. Rev. D {\bf 54}, 3830 (1996).
\bibitem{Henn}M. Henneaux, C. Mart\'{i}nez, R. Troncoso and J. Zanelli, Phys. Rev. D {\bf 65}, 104007 (2002).
\bibitem{Has1}M. Horta\c{c}su, H.T. \"{O}z\c{c}elik and B. Yap{\i}\c{s}kan, Gen. Rel. Grav. {\bf 35}, 1209 (2003).
\bibitem{Hasp}M. Hasanpour, F. Loran and H. Razaghian, Nucl. Phys. B {\bf 867}, 483 (2013).
\bibitem{Schmidt}H.J. Schmidt and D. Singleton, Phys. Lett. B {\bf 721}, 294 (2013).
\bibitem{Gar}A. Garc\'{i}a , F.W. Hehl, C. Heinicke and A. Mac\'{i}as, Phys. Rev. D {\bf 67}, 124016 (2003).
\bibitem{Miel}E.W.  Mielke  and A.A.R. Maggiolo,  Phys. Rev. D {\bf 68}, 104026 (2003).
\bibitem{Bla} M. Blagojevi\'{c}  and B. Cvetkovi\'{c},  Phys. Rev. D {\bf 78}, 0444036 (2008).
\bibitem{Bla1}M. Blagojevi\'{c} and B. Cvetkovi\'{c},  Phys. Rev. D {\bf 85}, 104003 (2012).
\bibitem{Bla2}M. Blagojevi\'{c}  and B. Cvetkovi\'{c}, Phys. Rev. D {\bf 88}, 104032 (2013).
\bibitem{Bla3}M. Blagojevi\'{c}, B. Cvetkovi\'{c} and M. Vasili\'{c}, Phys. Rev. D {\bf 88}, 101501 (2013).
\bibitem{Hehl}F.W. Hehl, Gen. Rel. Grav. {\bf 5}, 491 (1974).
\bibitem{Niko}J.N. Poplawski, arXiv:gr-qc/0911.0334.
\bibitem {Krec}V. Krechet and D.V. Sadovnikov, Russian Physics Journal {\bf 40}, 492 (1997).
\bibitem{Hojman} S. Hojman, M. Rosenbaum, M. P. Ryan, and L. C. Shepley, Phys. Rev. D {\bf 17}, 3141-3146
(1978). 
\bibitem{Spi}M. Spivak, {\it A Comprehensive Introduction to Differential
Geometry}, vol~2 3th ed. (Publish or Perish, Inc., Texas 1999).
\bibitem{Tunc} H.T. \"{O}z\c{c}elik, {\it PhD Thesis} (Y{\i}ld{\i}z Technical University, 2016).
\bibitem{Birrell} N.D. Birrell and P.C.W. Davies, {\it Quantum Fields in Curved Space}, p. 44, eq. 3.27, ( Cambridge University Press 1982).
\bibitem{Bunn}J.C. Baez and E.F. Bunn, Amer. Jour. Phys. {\bf 73}, 644-652 (2005), arXiv:gr-qc/0103044v6.
\bibitem{Neto} J.B. Fonseca-Neto, C. Romero and S.P.G. Martinez, Gen. Rel. Grav. {\bf 45}, 1579-1601 (2013), arXiv:gr-qc/1211.1557.

\end{thebibliography}
\end{document}